\documentclass[12pt]{article}
\usepackage{amsmath,epsfig,graphicx,amssymb,mathrsfs}
\usepackage[margin=0.7in]{geometry}
\newcommand{\beq}{\begin{equation}}
\newcommand{\eeq}{\end{equation}}
\newcommand{\ben}{\begin{eqnarray}}
\newcommand{\een}{\end{eqnarray}}
\newcommand{\bes}{\begin{subequations}}
\newcommand{\ees}{\end{subequations}}
\newcommand{\bFig}{\begin{figure}}
\newcommand{\eFig}{\end{figure}}

\date{}
\begin{document}
\title{Classical Electrodynamics in a Unified Theory}
\author{Partha Ghose\footnote{partha.ghose@gmail.com} \\
The National Academy of Sciences, India,\\ 5 Lajpatrai Road, Allahabad 211002, India\\and\\
Anirban Mukherjee\footnote{mukherjee.anirban.anirban@gmail.com},\\Indian Institute of Science Education \& Research,\\Mohanpur Campus, West Bengal 741252.}
\maketitle
\begin{abstract}
Some consequences of a fully classical unified theory of gravity and electromagnetism are worked out for the electromagnetic sector such as the occurrence of classical light beams with spin and orbital angular momenta that are topologically quantized in units of $q_e q_m=\sigma$, independent of the beam size. Empirical fits require $\sigma = \hbar$. The theory also predicts a generalized coherency matrix whose consequences are testable.
\end{abstract}
\section{Introduction}
In a recent paper \cite{pg} a unified classical theory of gravity and electromagnetism has been proposed which is based on a $U_4$ manifold with nonsymmetric connections $\Gamma^\lambda_{\mu\nu}$ and a nonsymmetric metric tensor $g^{\mu\nu}$ which can be split into symmetric and antisymmetric parts as follow:
\ben
s^{\mu\nu} &=& \frac{1}{2}\sqrt{-g}\left(g^{\mu\nu} + g^{\nu\mu}\right) =\frac{1}{2}\sqrt{-g}g^{(\mu\nu)},\\
a^{\mu\nu} &=& \frac{1}{2}\sqrt{-g}\left(g^{\mu\nu} - g^{\nu\mu}\right) = \frac{1}{2} \sqrt{-g} g^{[\mu\nu]},\\
\Gamma^\lambda_{\mu\nu} &=& \Gamma^\lambda_{(\mu\nu)} + \Gamma^{\,\,\,\,\,\,\,\,\,\,\lambda}_{[\mu\nu]},\\
\Gamma^\lambda_{(\mu\nu)} &=& \frac{1}{2}\left(\Gamma^\lambda_{\mu\nu}  + \Gamma^\lambda_{\nu\mu}\right),\\
\Gamma^{\lambda}_{\,\,\,\,[\mu\nu]} &=& \frac{1}{2}\left(\Gamma^\lambda_{\,\,\mu\nu}  - \Gamma^\lambda_{\,\,\nu\mu}\right).
\een 
$\Gamma^{\lambda}_{\,\,\,\,[\mu\nu]}$ is the Cartan torsion and $\Gamma_\mu = \Gamma^\lambda_{[\mu\lambda]}$ is the torsion pseudovector.
It has many implications for classical optics. 

The fundamental equation for electrodynamics in this theory is
\ben
\tilde{F}^{\mu\nu}_{;\,\nu} &=& l^\mu,\label{maxwell}
\een
which implies
\beq
l^\mu_{,\,\mu} = 0.\label{l}
\eeq
Defining the fields
\beq
\tilde{F}^{0i} = - B^{i},\,\,\,\,\,\,\tilde{F}^{ij} = \frac{1}{c}\epsilon^{ijk}E_k, 
\eeq 
and $l^\mu = (-\mu_0\rho_m, -{\bf j}_m/c\epsilon_0)$, one gets from (\ref{maxwell}) the equations
\ben
{\bf \nabla} \times {\bf E} + \frac{\partial {\bf B}}{\partial t} &=& -\frac{1}{\epsilon_0} \bf{j}_m,\,\,\,\,\,\,\bf{\nabla}. {\bf{B}} = \mu_0\rho_m.\label{ma2}
\een
Hence, {\em there is a magnetic current density in the theory proportional to $\Gamma^\mu$}, and $\tilde{F}_{\mu\nu}$ can be interpreted as the dual of the electromagnetic field. 

These equations can also be written in the form
\ben
F_{\mu\nu,\,\lambda} + F_{\nu\lambda,\,\mu} + F_{\lambda\mu,\,\nu} &=& \epsilon_{\mu\nu\lambda\rho}l^\rho \label{B1},\\
\tilde{F}^{\mu\nu} = \frac{1}{2}\epsilon^{\mu\nu\lambda\rho}F_{\lambda\rho},\label{tild} 
\een
where $\epsilon_{\mu\nu\lambda\rho}$ is the Levi Civita tensor density with components $\pm 1$. 

The electric current density $j^\mu$ in this theory is given by
\ben
j^\mu &=& \frac{1}{3!}\epsilon^{\mu\nu\lambda\rho}\left(\tilde{F}_{\nu\lambda,\,\rho} + \tilde{F}_{\lambda\rho,\,\nu} + \tilde{F}_{\rho\nu,\,\lambda}\right)\label{current}\\
&=& F^{\mu\nu}_{\,\,\,\,,\,\nu}.\label{maxwell3}
\een
Hence,
\beq
j^\mu_{,\,\mu} = 0.\label{j}
\eeq
Using the definitions $j^\mu = (-\rho_q/c\epsilon_0, -\mu_0{\bf j}_q)$ and 
\beq
F^{0i} = -\frac{E^i}{c},\,\,\,\,\,\, F^{ij} = - \epsilon^{ijk} B_k, 
\eeq
one then obtains
\ben
{\bf{\nabla} \times {\bf{B}}} - \frac{1}{c^2}\frac{\partial {{\bf{E}}}}{\partial t} &=& \mu_0 {\bf j}_q,\,\,\,\,\,\,\bf{\nabla}. \bf{E} = \frac{1}{\epsilon_0}\rho_q. \label{ma1}
\een
These equations can also be written in the form
\ben
\tilde{F}_{\mu\nu,\,\lambda} + \tilde{F}_{\nu\lambda,\,\mu} + \tilde{F}_{\lambda\mu,\,\nu} = \epsilon_{\mu\nu\lambda\rho}j^\rho. \label{B2}
\een

\section{Duality}
 
The equations (\ref{ma2}) and (\ref{ma1}) are together invariant under the generalized (Heaviside) duality transformations \cite{heaviside}
\ben
{\bf E} &\rightarrow& c{\bf B},\nonumber\\
c{\bf B} &\rightarrow& -{\bf E},\nonumber\\
(\rho_q, {\bf j}_q) &\rightarrow& (\rho_m, {\bf j}_m),\nonumber\\
(\rho_m, {\bf j}_m) &\rightarrow& (-\rho_q, -{\bf j}_q).\label{heaviside}
\een
This symmetry is therefore a {\em consequence} of the unified theory. In fact, using {\em split-complex} tensors, equations (\ref{maxwell}) and (\ref{maxwell3}) can be combined into a single equation 
\ben
\partial_\nu H^{\mu\nu} &=& J^\mu,\label{R}\\
H^{\mu\nu} &=& \frac{1}{\sqrt{2}}\left(F^{\mu\nu} + j \tilde{F}^{\mu\nu}\right),\\
J^\mu &=& \frac{1}{\sqrt{2}}\left(j^\mu + j l^\mu\right), 
\een
where $j^2 =1, \, j\neq \pm 1$.
Eqn. (\ref{R}) is invariant under the continuous $U(1)$ transformations
\cite{larmor}
\ben
H^{\mu\nu} &\rightarrow& e^{i\theta}H^{\mu\nu},\\
J^\mu &\rightarrow& e^{i\theta} J^\mu\,\,\,\,\,\,\theta \in [0,\, 2\pi]
\een
provided $\theta$ is a global phase factor.
Consequently, the moduli
\ben
||H|| &=& H^{\mu\nu*}H_{\mu\nu} = \frac{1}{2}\left(F^{\mu\nu}F_{\mu\nu} - \tilde{F}^{\mu\nu}\tilde{F}_{\mu\nu}\right) = F^{\mu\nu}F_{\mu\nu}= - 2\left(\frac{E^2}{c^2} - B^2\right),\\
||J|| &=& J^{\mu*}J_\mu = \frac{1}{2}\left(j^\mu j_\mu - l^\mu l_\mu\right)
\een 
are invariants under these phase transformations.

The following are invariants under general coordinate transformations:
\ben
F^{\mu\nu}F_{\mu\nu} &=& - 2\left(\frac{E^2}{c^2} - B^2\right),\\
\tilde{F}^{\mu\nu}\tilde{F}_{\mu\nu} &=& - 2 \left(B^2 - \frac{E^2}{c^2}\right),\\
F^{\mu\nu}\tilde{F}_{\mu\nu} &=& - \frac{4}{c} {\bf E}.{\bf B}.\label{eb}
\een
However, they change sign under the Heaviside transformation.

\section{Singularities}

One can also form ordinary complex combinations
\beq
Z^{\mu\nu} = a F^{\mu\nu} + ib \tilde{F}^{\mu\nu}
\eeq
with $a^2 + b^2 = 1$. For $a = b = 1/\sqrt{2}$ the modulus (norm) vanishes:
\beq
||Z|| = Z^{\mu\nu*}Z_{\mu\nu} =\frac{1}{2}\left(F^{\mu\nu}F_{\mu\nu} + \tilde{F}^{\mu\nu}\tilde{F}_{\mu\nu}\right) = 0. \label{sing}
\eeq
Writing $Z^{\mu\nu} = \rho^{\mu\nu}{\rm exp}(i\phi)$, we have $\rho^{\mu\nu}\rho_{\mu\nu} = 0$. Hence, there is a phase singularity for such fields \cite{dennis}. Notice that the components of such a field are $\sqrt{\mu}_0$ times the Riemann-Silberstein vector $F_i = \sqrt{\frac{\epsilon_0}{2}}(E_i + icB_i)$ \cite{rs}.

If one introduces potentials through the relations
\ben
F_{\mu\nu} &=& \partial_\mu A_\nu - \partial_\nu A_\mu,
\een
they turn out to be singular in this theory. A static magnetic field due to a charge $g$ is given by
\beq
\vec{B} = g \frac{\bf r}{r^3} = {\nabla} \times {\bf A},
\eeq 
but this is contradicted by eqn. (\ref{ma2}). The solution lies in the famous Dirac potential \cite{dirac} which can be written in spherical polar coordinates as
\beq
{\bf A}_\phi = \frac{g}{r}{\rm tan}\frac{\theta}{2}\hat{\phi},\,\,\,\,A_r = A_\theta = 0
\eeq
with the solution 
\beq
{\bf B}_r = g \frac{\bf r}{r^3},\,\,\,\,{\bf B}_\phi = {\bf B}_\theta =0.  
\eeq
This potential is clearly singular along the negative $z$ axis characterized by $\theta = \pi$. This is the famous Dirac string. Since the theory under consideration is a field theory, the potential will be non-holomorphic in general, and there will be a magnetic monopole density at the origin. 

Let us recall that the relations between the fields and potentials in standard electrodynamics are
\ben
{\bf B} &=& {\bf \nabla} \times {\bf A},\\
{\bf E} &=& - {\bf \nabla}\phi + \frac{1}{c}\frac{\partial {\bf A}}{\partial t}.
\een
These are not symmetric under the Heaviside transformations (\ref{heaviside}). To have this symmetry at the level of the potentials it is necessary to introduce pseudovector potentials $\tilde{A}_\mu$ defined by
\beq
\tilde{F}_{\mu\nu} = \partial_\mu \tilde{A}_\nu - \partial_\nu \tilde{A}_\mu 
\eeq and the relations
\ben
{\bf E} &=& {\bf \nabla} \times \tilde{{\bf A}},\label{E2}\\
{\bf B} &=& - {\bf \nabla}\tilde{\phi} + \frac{1}{c}\frac{\partial \tilde{{\bf A}}}{\partial t}.
\een
However, the first of these relations is contradicted by eqn. (\ref{ma1}). Hence, like $A_\mu$, $\tilde{A}_\mu$ must also be a non-holomorphic function.

\subsection{Circulation and Angular Momentum Quantization}

It has been shown in \cite{pg} that $\Gamma_\mu$ has two properties: it is the source of a magnetic current $l^\mu$ whose time component is a magnetic monopole density $\rho_m$, and it is also curlfree:
\beq
\Gamma_{\mu,\nu} - \Gamma_{\nu,\mu} = 0.\label{curlfree}
\eeq
Consequently,
\ben
l_{\mu,\nu} - l_{\nu,\mu} &=& 0,\label{curlfree2}\\
j_{\mu,\nu} - j_{\nu,\mu} &=& 0.\label{curlfree3}
\een
It follows from these equations that
\beq
\Box l_\mu = \Box j_\mu = 0,
\eeq
which can be expanded in the forms
\begin{eqnarray}
\nabla\rho_{m}- \frac{\partial\mathbf{j}_{m}}{\partial t}&=&0,\,\,\nabla\times\mathbf{j}_{m}=0,\nonumber\\
\nabla\rho_{q} -\frac{1}{c^2}\frac{\partial\mathbf{j}_{q}}{\partial t}&=&0,\,\,\nabla\times\mathbf{j}_{q}=0,\nonumber\\
\nabla\cdot\mathbf{j}_{m}+ \frac{1}{c^2}\frac{\partial\rho_{m}}{\partial t}&=&0,\,\,\nabla\cdot\mathbf{j}_{q}+ \frac{\partial\rho_{q}}{\partial t}=0,\nonumber\\
\left[\nabla^{2}-\frac{1}{c^2}\frac{\partial^{2}}{\partial t^{2}}\right]l_{\mu}&=&0,\,\,\left[\nabla^{2}-\frac{1}{c^2}\frac{\partial^{2}}{\partial t^{2}}\right]j_{\mu}=0. \label{equation set constraints}
\end{eqnarray}
By taking the curl of the Maxwell equations (\ref{ma2}) and making use of (\ref{ma1}) and the above results, one gets
\begin{eqnarray}
\nabla\times\nabla\times\mathbf{E}+\frac{\partial}{\partial t}\nabla\times\mathbf{B}&=&\nabla\times\mathbf{j}_{m},\nonumber\\
{\rm or}\,\,\nabla(\nabla\cdot\mathbf{E})-\nabla^{2}\mathbf{E}+ \frac{1}{c^2}
\frac{\partial^{2}\mathbf{E}}{\partial t^{2}}+\mu_0\frac{\partial\mathbf{j}_{q}}{\partial t}&=&0,
\nonumber\\
{\rm or}\,\, \left[\nabla^{2}-\frac{1}{c^2}\frac{\partial^{2}}{\partial t^{2}}\right]\mathbf{E}&=&0 \label{freeE}.
\end{eqnarray}
Similarly, one gets
\begin{eqnarray}
\left[\nabla^{2}-\frac{1}{c^2}\frac{\partial^{2}}{\partial t^{2}}\right]\mathbf{B}&=&0. \label{freeB}
\end{eqnarray}
The solutions to these equations are given by
\begin{eqnarray}
\mathbf{E}(\mathbf{r},t)&=&\frac{1}{L^{3/2}}\sum_{\mathbf{k},\omega_k}\left[\mathbf{c}_{\mathbf{k}\omega_k}
e^{i(\mathbf{k}\cdot\mathbf{r}-\omega_k t)}+\mathbf{c}^{*}_{\mathbf{-k}\omega_k}
e^{i(\mathbf{k}\cdot\mathbf{r}+\omega_k t)}\right]\nonumber\\
\mathbf{B}(\mathbf{r},t)&=&\frac{1}{L^{3/2}}\sum_{\mathbf{k},\omega_k}\left[\mathbf{d}_{\mathbf{k}\omega_k}
e^{i(\mathbf{k}\cdot\mathbf{r}-\omega_k t)}+\mathbf{d}^{*}_{\mathbf{-k}\omega_k}
e^{i(\mathbf{k}\cdot\mathbf{r}+\omega_k t)}\right]
\end{eqnarray}
where the limit $L \rightarrow \infty$, $\sum_{\mathbf{k},\omega_k} \rightarrow \int d^3 k/(2\pi)^{3/2}$ is assumed.

Now, because of the conditions (\ref{curlfree2}) and (\ref{curlfree3}), the electric and magnetic current densities can be written as
\begin{eqnarray}
j_{\mu}&=&\partial_{\mu}\Phi_{1},\nonumber\\
l_{\mu}&=&\partial_{\mu}\Phi_{2}.
\end{eqnarray} 
Hence, it follows from the conservation eqns. (\ref{l}) and (\ref{j}) that $\Box \Phi_{1,2} =0$, whose soutions can be written as
\begin{eqnarray}
\Phi_{1,2}=\sum_{\mathbf{k},\omega_k}\left[\phi_{1,2\mathbf{k}\omega_k}
e^{i(\mathbf{k}\cdot\mathbf{r}-\omega_k t)}+\phi^{*}_{1,2-\mathbf{k}\omega_k}
e^{i(\mathbf{k}\cdot\mathbf{r}+\omega_k t)}\right].
\end{eqnarray} 
Feeding these solutions into the Maxwell equations (\ref{ma2}) and (\ref{ma1}), we have
\begin{eqnarray}
\mathbf{k}\cdot\mathbf{c}_{\mathbf{k}\omega_k}&=&
-\omega_k\phi_{1\mathbf{k}\omega_k},\,\mathbf{k}\cdot\mathbf{d}_{\mathbf{k}\omega_k}=
-\omega_k\phi_{2\mathbf{k}\omega_k}\nonumber\\
\mathbf{k}\times\mathbf{c}_{\mathbf{k}\omega_k}&=&
\omega_k\mathbf{d}_{\mathbf{k}\omega_k}+\mathbf{k}\phi_{2\mathbf{k}\omega_k},\, \mathbf{k}\times\mathbf{d}_{\mathbf{k}\omega_k}=
-\frac{\omega_k}{c^2}\mathbf{c}_{\mathbf{k}\omega_k}+
\mathbf{k}\phi_{1\mathbf{k}\omega_k}\label{equations_maxwell}
\end{eqnarray}

The corresponding equations in standard Maxwell theory in free space (i.e. in the absence of $l^\mu$ and $j^\mu$) are given by
\begin{eqnarray}
\mathbf{k}\cdot{\mathbf{c}}_{\mathbf{k}\omega_k}&=&0,\, \mathbf{k}\cdot{\mathbf{d}}_{\mathbf{k}\omega_k}=0,\nonumber\\
\mathbf{k}\times\mathbf{c}_{\mathbf{k}\omega_k}&=&
\omega\mathbf{d}_{\mathbf{k}\omega_k},\,\mathbf{k}\times\mathbf{d}_{\mathbf{k}\omega_k}= -\frac{\omega_k}{c^2}\mathbf{c}_{\mathbf{k}\omega_k}.
\end{eqnarray} 
A single mode electromagnetic field that satisfies these free space equations is characterized by
\begin{eqnarray}
\mathbf{E}&=& |{\bf E}|\hat{r} = \left(E_{0}e^{i(k_{z}z-\omega_k t)}+E^{*}_{0}e^{-i(k_{z}z-\omega_k t)}\right)\hat{r},\\
\mathbf{B}&=& |{\bf B}|\hat{\phi} =\left(B_{0}e^{i(k_{z}z-\omega_k t)}+B^{*}_{0}e^{-i(k_{z}z-\omega_k t)}\right)\hat{\phi},\nonumber\\
&=&\frac{1}{c}|{\bf E}|\hat{\phi},\\
\mathbf{k}&=&k_{z}\hat{z},\, E_{0}\hat{r} =\mathbf{c}_{\mathbf{k}\omega_k},\,\frac{1}{c}E_{0}\hat{\phi} =\mathbf{d}_{\mathbf{k}\omega_k},
\end{eqnarray}
where $\omega_k = c |{\bf k}|$. Hence the Poynting vector for this case is
\begin{eqnarray}
\mathbf{S}&=&\frac{1}{\mu_0}\mathbf{E}\times\mathbf{B}=\frac{1}{\mu_0}(|\mathbf{E}||\mathbf{B}|)\hat{z}.
\end{eqnarray}
Consequently,
\begin{eqnarray}
\oint\mathbf{S}\cdot \mathbf{dl}&=&0,
\end{eqnarray}
and ${\bf S}$ has no winding number configurations. Further, $\mathbf{E}\cdot\mathbf{B}=0$.

Let us now consider a wave-vector propagating in the $\hat{z}$ direction $\mathbf{k}=k_{z}\hat{z}$ corresponding to the solutions (\ref{equations_maxwell}):
\begin{eqnarray}
\mathbf{E}&=&|{\bf E}|\hat{r} +\left(\phi_{1}e^{i(k_{z}z-\omega_k t)}+\phi_{1}^{*}e^{-i(k_{z}z-\omega_k t)}\right)\hat{z},\nonumber\\
\mathbf{B}&=& |{\bf B}|\hat{\phi} +\left(\phi_{2}e^{i(k_{z}-\omega_k t)}+\phi^{*}_{2}e^{-i(k_{z}-\omega_k t)}\right)\hat{z}.\label{sol}
\end{eqnarray}
The Poynting vector in this case is given by
\begin{eqnarray}
\mu_0\mathbf{S}&=& \mathbf{E}\times\mathbf{B} \nonumber\\
		  &=&(|\mathbf{E}||\mathbf{B}|)\hat{z} \\
		  &+& |\mathbf{E}|\left(\phi_{2}e^{i(k_{z}z-\omega_k t)}+\phi_{2}^{*}e^{-i(k_{z}z-\omega_k t)}\right)\hat{\phi}	\nonumber\\
		  &-&|\mathbf{B}|\left(\phi_{1}e^{i(k_{z}z-\omega_k t)}+\phi_{1}^{*}e^{-i(k_{z}z-\omega_k t)}\right)\hat{r}.
			\end{eqnarray}
The time-averaged vector $\langle \mathbf{S}\rangle$ can be written as
\begin{eqnarray}
\langle\mathbf{S}\rangle &=& \frac{1}{\mu_0}\left[|E_{0}|^{2}\hat{z} + \left(E_{0}\phi_{2}^{*}+ E^{*}_{0}\phi_{2}\right)\hat{\phi}- (E_{0}\phi^{*}_{1} + E_{0}^{*}\phi_{1})\hat{r}\right]\nonumber\\
&\equiv& 2\langle S_z\rangle\hat{z} + \langle S_\phi\rangle\hat{\phi} - \langle S_r\rangle\hat{r}.\label{S}
\end{eqnarray}
Hence, $\nabla\times\langle\mathbf{S}\rangle =0$. However, remembering that the momentum density of the electromagnetic field is ${\bf P} = {\bf S}/c^2$, we have 
\begin{eqnarray}
\oint \mathbf{P}\cdot d\mathbf{l} =\frac{1}{c^2} \oint \mathbf{S}\cdot d\mathbf{l} =\frac{1}{c^2} \oint \langle S\rangle_{\phi}rd\phi = \pm 2n\pi\sigma
\end{eqnarray}
where $n$ is a winding number, $\sigma$ is the unit of angular momentum in the theory (to be determined in the next section), and r ranges from $[0,R]$, $R$ being the beam radius. 
Therefore, there is a topological constraint induced by the winding number, namely
\begin{eqnarray}
\langle S\rangle_{\phi}= \frac{1}{\mu_0} \left(E_{0}\phi_{2}^{*}+
E^{*}_{0}\phi_{2}\right) =\frac{c^2 n\sigma}{r}.
\end{eqnarray}
Consequently, the angular momentum associated with this beam is given by
\begin{eqnarray}
\mathbf{L}=r\hat{r}\times \mathbf{P}= r\hat{r}\times \frac{n\sigma}{r}\hat{\phi}=n\sigma\hat{z}.\label{L}
\end{eqnarray}

Now, it can be shown that this total angular momentum can be split into an orbital part and a spin part in a gauge invariant way by using the Helmholtz decomposition of $\mathbf{E}$ and $\mathbf{B}$ \cite{stewart}. The Helmholtz decomposition gives 
\ben
\mathbf{E}(\mathbf{x},t) &=& - \mathbf{\nabla}_x f(\mathbf{x},t) + \mathbf{\nabla}_x\times \mathbf{F}(\mathbf{x},t)
\een
with
\ben
f(\mathbf{x},t) &=& \int d^3 y \frac{\mathbf{\nabla}_y\cdot\mathbf{E}(\mathbf{y},t)}{|\mathbf{x}-\mathbf{y}|}\nonumber\\
&=&\int d^3 y \frac{\rho_q(\mathbf{y},t)}{\epsilon_0|\mathbf{x}-\mathbf{y}|},\\
\mathbf{F}(\mathbf{x},t) &=&\int d^3 y \frac{\mathbf{\nabla}_y\times\mathbf{E}(\mathbf{y},t)}{|\mathbf{x}-\mathbf{y}|},\nonumber\\
&=&-\int d^3 y \frac{\frac{\partial}{\partial t}\mathbf{B}(\mathbf{y},t)+\frac{1}{\epsilon_0}\mathbf{j}_m(\mathbf{y},t)}{|\mathbf{x}-\mathbf{y}|}.
\een
Using these results and the vector identity
\beq
\left(\mathbf{\nabla}_x \times \mathbf{F}\right) \times \mathbf{B} = \left(\mathbf{B}\cdot\mathbf{\nabla}_x\right)\mathbf{F} - \sum^3_{r=1} B^r\mathbf{\nabla}_x F^r,
\eeq 
the angular momentum can be written as 
\ben
\mathbf{L}(t) &=& \frac{1}{\mu_0 c^2}\int d^3 x\, \mathbf{x}\times [\mathbf{E}(\mathbf{x},t)\times \mathbf{B}(\mathbf{x},t)]\nonumber\\
&=& \mathbf{L}_s + \mathbf{L}_o
\een
with
\ben
\mathbf{L}_s &=& \int d^3 x \mathbf{F}\times \mathbf{B} = -\int d^3 x \int d^3 y \frac{\partial_t\mathbf{B}(\mathbf{y},t) + \frac{1}{\epsilon_0}\mathbf{j}_m(\mathbf{y},t)}{|\mathbf{x}-\mathbf{y}|}\times \mathbf{B}(\mathbf{x},t)\\
\mathbf{L}_o &=& \int d^3 x \int d^3 y \left[\mathbf{B}(\mathbf{x},t)\cdot \frac{\partial \mathbf{B}(\mathbf{y},t)}{\partial t}\right]\frac{\mathbf{x}\times \mathbf{y}}{|\mathbf{x} - \mathbf{y}|^3}\nonumber\\ &-& \int d^3 x \rho_m(\mathbf{x},t)\int d^3 y \frac{\left(\partial_t \mathbf{B}(\mathbf{y},t) + \frac{1}{\epsilon_0}\mathbf{j}_m(\mathbf{y},t)\right)\times \mathbf{x}}{|\mathbf{x}-\mathbf{y}|}
\een
These equations reduce to the ones given in Ref. \cite{stewart} for $\rho_m = \mathbf{j}_m=0$.
By writing the first term in the component form
\beq
L_s^i = -\epsilon_{ijk}\int d^3 x B^j F^k = (s_j)_{ik}\int d^3 x B^j F^k
\eeq
where $s_i$ are $3\times 3$ rotation matrices with the commutation relations
\beq
\left[s_i,\, s_j \right] = \epsilon_{ijk}s_k
\eeq
which describe the rotational aspects of the fields, it becomes clear that it is the spin part of the total angular momentum. The second term $\mathbf{L}_o$ is the orbital angular momentum part.

Standard classical optics permits plane wave solutions of arbitrary extent (at least ideally) with the Poynting vector in the direction of propagation. Hence, they do not have any angular momentum in the direction of propagation. Yet, experiments have clearly shown that circularly polarized classical light does carry angular momentum \cite{raman, beth}. It is now widely believed that the angular momenta of light beams arise from their finite size (edge effects) and the paraxial approximation which describes laser beams fairly well \cite{padget}. However, such approximations cannot explain why the observed angular momenta of {\em classical} light are quantized in units of $\hbar$. It is therefore further assumed that the quantization is due to the average spin angular momentum per photon in the beams. 

We have found that in spite of the inhomogeneous Maxwell equations (\ref{ma2}) and (\ref{ma1}), the electric and magnetic fields satisfy the free field equations (\ref{freeE}) and (\ref{freeB}) as a consequence of the curlfree condition (\ref{curlfree}). However, the solutions (\ref{sol}) to the free field equations show that they are not plane waves normal to the direction of propagation $\hat{z}$---they have components along the $\hat{z}$ direction. This is why the Poynting vector picks up a component in the azimuthal direction $\hat{\phi}$, and that leads, because of the Dirac strings attached to magnetic monopoles, to a topologically quantized total angular momentum in the $\hat{z}$ direction. Consequently, {\em the Poynting vector follows a helical trajectory about the propagation direction}. Unlike in standard classical optics, this is an inherent feature of {\em all} optical waves predicted by the theory. However, it may be difficult to observe such behaviour except when optical waves pass through apertures or encounter obstacles. States with winding number (or topological charge) $n=1$ are circularly polarized with helicity $\pm 1$, and the orbital angular momentum of such states must be zero. States with $n\geq 2$ carry both helicity and orbital angular momentum. Only states with the winding number $n=0$ have no total angular momentum and correspond to plane polarized light. Plane polarized light must therefore be an equal superposition of $+1$ and $-1$ helicity states.  No quantum theory has been used at any stage to arrive at these results. 

In standard classical optics the orbital angular momentum of classical light beams arise only in the paraxial approximation when the Poynting vector is not parallel to the beam axis and follows a spiral trajectory about that axis. Notice that the total angular momentum of a light beam according to the unified theory is given by eqn. (\ref{L}) which is independent of $r$, the beam size, and is entirely topological in character. A crucial test to distinguish between standard classical optics and the optics implied by the unified theory would therefore be to see if the spin angular momentum of circularly polarized light as well as the orbital angular momentum of vortex beams are independent of the beam size.

\section{Charge Quantization}
Let us consider a system consisting of an electric charge $q_{e}$ and a Dirac string carrying magnetic charge $q_{m}$ as shown in the figure. The charges are defined by
\beq
q_e = \frac{\rho_q}{\bar{n}},\,\,\,\,q_m = \frac{\rho_m}{\bar{m}}
\eeq
where $\bar{n}$ and $\bar{m}$ are number densities of electric and magnetic charges.
\begin{figure}
\includegraphics[scale=0.5]{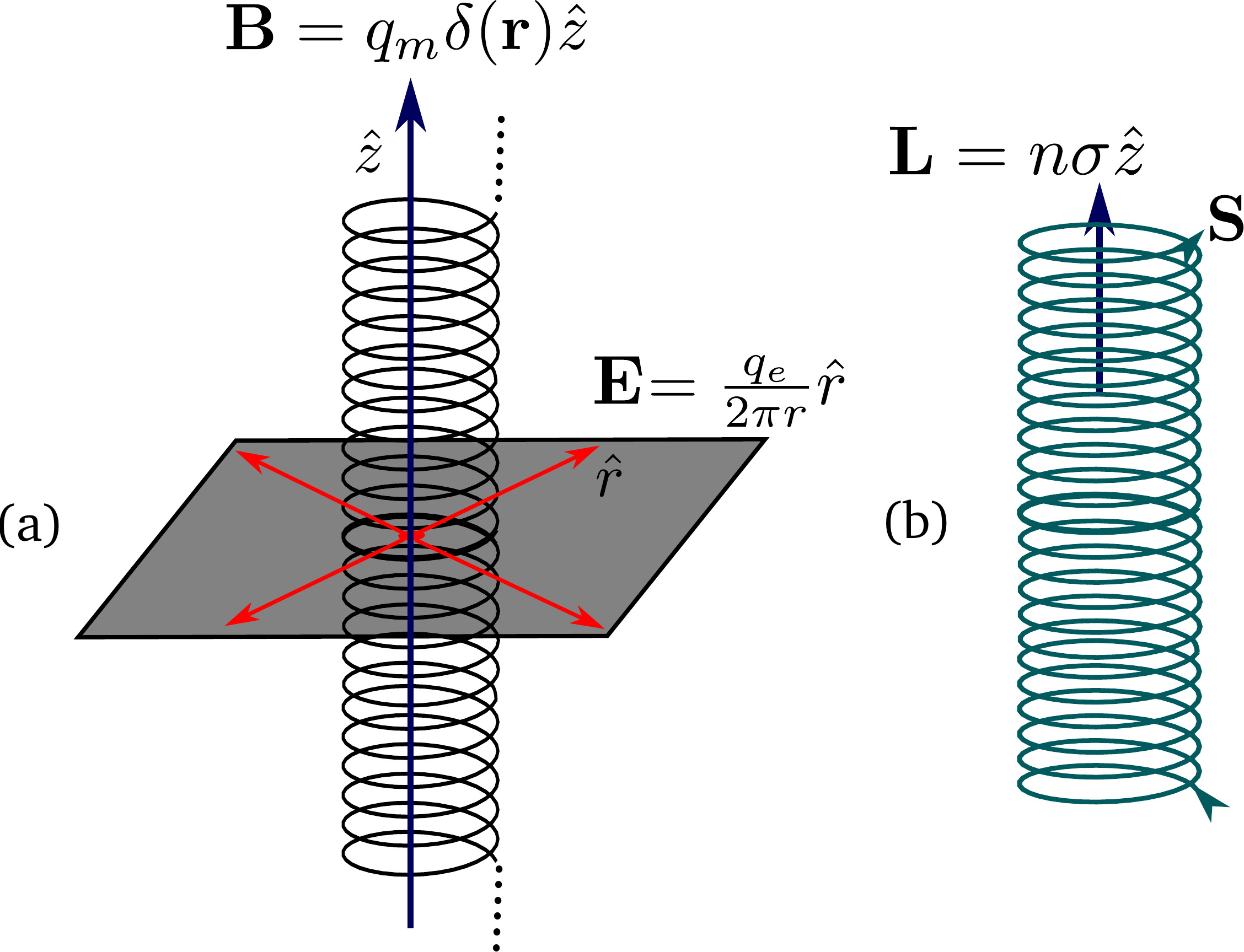}
\caption{Fig (a) represents a 2D radial electric field due to a charge $q_e$ and a Dirac string with a magnetic charge $q_m$ at one end. Fig (b) shows the Poynting vector $\mathbf{S}$ with an azimuthal component $S_{\phi}\hat{\phi}$, propogating in the z direction} \label{helix}
\end{figure}
The 2D electric field due to the electric charge and the magnetic field due to the semi-infinite, infinitely thin solenoid (the Dirac string) are given by
\begin{eqnarray}
\mathbf{E}&=&\frac{q_{e}}{2\pi\epsilon_0 r}\hat{r}, \quad \hat{r}=\cos\phi\hat{x}+ \sin\phi\hat{y},\nonumber\\
\mathbf{B}&=&q_{m}\delta(r)\hat{z}.
\end{eqnarray}
The Poynting vector is given given by
\begin{eqnarray}
\mathbf{S}=\frac{1}{\mu_0}\mathbf{E}\times\mathbf{B}
=\frac{q_{e}q_{m}}{2\pi\epsilon_0\mu_0 r}\delta(r)\hat{\phi}.
\end{eqnarray}
Therefore, the angular momentum of the electromagnetic field created by the electric charge-magnetic monopole pair is 
\begin{eqnarray}
\mathbf{L}_s= \frac{1}{c^2}\int rdrd\theta\, \mathbf{r}\times\mathbf{S} =q_{e}q_{m}\hat{z}.
\end{eqnarray}
This is clearly a spin angular momentum, and therefore using $n=1$ in (\ref{L}), we get
\begin{eqnarray}
q_{e}q_{m}=\sigma.
\end{eqnarray}
This determines $\sigma$, the unit of angular momentum in the theory. It shows that the quantized angular momenta of electromagnetic fields have their origin in the magnetic monopole $q_m$, i.e. $\Gamma_0$, the time component of the curlfree torsion pseudovector $\Gamma_\mu$. Furthermore, since electric charge is quantized in units of $e$, i.e. $q_e = e$, $q_m = \sigma/e$ must also be quantized. Empirical fits require that $\sigma = 1.054571800(13) \times 10^{-13} J.s = \hbar$. Hence, Planck's constant emerges in the unified theory as a result of topological quantization of angular momentum. Though a remarkable result, quantum theory does not follow from it alone.

\section{Polarization}
 
The standard coherency matrix $J = \langle EE^*\rangle$ (defined in the $(x,y)$ plane) is not a Heaviside invariant. However, the generalized coherency matrix
\beq
F = \langle EE^*{\rm sin}^2\,\phi + c^2 BB^*{\rm cos}^2\,\phi\rangle
\eeq
is Heaviside invariant for $\phi = \pi/4$ and reduces to $J$ for $\phi = \pi/2$. For $\phi = \pi/4$ there is equal contribution from the ${\bf E}$ and ${\bf B}$ fields to $F$. Thus, $\phi$ is a free parameter in the theory related to Heaviside symmetry breaking.

One can define new Stokes parameters $S^B_\mu$ corresponding to the magnetic field and the state of polarization would be a point on a new Poincar\'{e} sphere whose axes are $S^B_i,\,i= x,y,z$. In general, one would need to consider both the Poincar\'{e} spheres in characterizing optical states for $\phi \neq 0,\, \pi/2$. Since the magnetic field is a pseudovector, the $B$ polarized states have opposite parity to the $E$ polarized states. Such polarization states are a definite prediction of the theory. 

$E$ polarization characteristics can be explained by adopting a simple classical model of electrons bound to fixed nuclei by an elastic force \cite{becker}. In this model the electron is driven into a damped oscillatory motion by the $E$ component of the incident wave. In order to produce $B$ polarization, magnetic charges should be driven to dipole oscillations by the $B$ component of the incident wave. 

In the unified theory the Lorentz force densities on the charge densities $\rho_q$ and $\rho_m$ due to external fields (generated by similar charge densities elsewhere) are given by
\ben
{\bf f}_q &=& \rho_q \left[ {\bf E} + {\bf v}\times {\bf B}\right],\\ 
{\bf f}_m &=& c\rho_m \left[{\bf B} - \frac{1}{c^2}{\bf v}\times {\bf E}\right]. 
\een
This shows that when electric charges are responsible for scattering, the $E$ component is dominant over the $B$ component for small velocities $v$. However, when magnetic charges are responsible fot scattering, the $B$ component will dominate over the $E$ comonent. The magnetic dipole moment is given by
\beq
{\mathbf{\mu}}({\bf r}) = \int \rho_m ({\bf r}_0)\left({\bf r} - {\bf r}_0 \right) d^3 {\bf r}_0.
\eeq

The unified theory does not necessarily predict directly observable magnetic monopoles. However, as we have seen in the last section, it predicts discrete spin and orbital angular momentum states in classical optics which would not be possible without them. Another indirect test would be to simulate a magnetic monopole with a Dirac string attached to it by producing sufficiently long nanowires of dielectric materials like silicon which exhibit magnetic dipole resonances \cite{nano}. The wire should consist of a line of magnetic dipoles with the near end carrying a magnetic monopole. The $\mathbf{B}$-component of an incident unpolarized electromagnetic wave will excite dipolar oscillations of the monopole, causing $B$ polarized electromagnetic waves to be emitted by it. The propagation of the dipolar oscillations of the monopole along the wire should be dampened so that the other end does not radiate. The detection of such $B$ polarized electromagnetic waves must also be done using dielectric nano resonators. The detection of such phenomena would not constitute a test of the unified theory, but would be inspired by magnetic monopoles, a necessary consequence of the unified theory, and may have important novel applications in technology.
\section{Conclusion}
On the basis of a unified theory of gravity and electromagnetism with a non-vanishing torsion pseudovector $\Gamma_\mu$ we have shown that classical light carries topologically quantized spin and orbital mementum. Hence, observations of optical states with quantized spin and orbital angular momenta are tell-tale signatures of magnetic monopoles and of unification. As is well known, magnetic monopoles also imply charge quantization. 

Furthermore, since the unified theory predicts toplogically quantized spin and angular momentum of classical light, it also predicts entanglement between the various modes of classical light and the violation of Bell-like inequalities \cite{ent, q} {\em independent of any approximation like the paraxial approximation}. Hence not only entanglement, quantized spin and angular momenta are also shared by classical and quantum systems. This requires a new interpretation of the classical-quantum boundary.  

\section{Acknowledgement}
PG gratefully acknowledges financial support from the National Academy of Sciences, India.

\end{document}